\documentclass[letterpaper, 10 pt, conference]{ieeeconf}  

\IEEEoverridecommandlockouts                              

\usepackage{graphics} 
\usepackage{epsfig} 

\usepackage{amsthm, amsmath, amssymb, stmaryrd, hyperref} 
\usepackage[noadjust]{cite} 

\newtheorem{corollary}{Corollary}

\newtheorem{lemma}{Lemma}

\newtheorem{proposition}{Proposition}
\newtheorem{theorem}{Theorem}

\theoremstyle{definition}
\newtheorem{definition}{Definition}
\newtheorem*{problem}{Problem Statement}
\newtheorem{assumption}{Assumption}

\DeclareMathOperator*{\argmin}{argmin}
\DeclareMathOperator*{\suchthat}{\mathrm{s.t.}}

\DeclareMathOperator*{\until}{\mathrm{U}}

\DeclareMathOperator*{\true}{\mathrm{True}}
\DeclareMathOperator*{\false}{\mathrm{False}}
\DeclareMathOperator{\testspace}{\mathcal{D}}

\DeclareMathOperator{\signalspace}{\mathcal{S}^{\mathbb{R}^n}_{t_f}}
\usepackage{amsmath}

\newcommand{\spacing}{\vspace{0.1 cm}}

\newcommand{\newidea}[1]{\noindent \textbf{#1}}

\newcommand{\simpledef}[2]{\begin{definition}
\label{#1}
#2
\end{definition}
\spacing
}

\newcommand{\simpleassumption}[2]{
\begin{assumption}
\label{#1}
#2
\end{assumption}
\spacing
}

\newcommand{\simpletheorem}[2]{
\begin{theorem}
\label{#1}
#2
\end{theorem}
\spacing
}

\newcommand{\simpleproposition}[2]{
\begin{proposition}
\label{#1}
#2
\end{proposition}
}

\newcommand{\probstate}[1]{
\begin{problem}
#1
\end{problem}
\spacing
}

\def\sq{\mathbin{{\strut\rule{1.25ex}{1.25ex}}}}
\renewenvironment{proof}{{\textbf{Proof:}}}{\hfill$\sq$}

\title{\LARGE \bf
Formal Verification of Safety Critical Autonomous Systems \\ via Bayesian Optimization
}

\author{Prithvi Akella, Ugo Rosolia, Andrew Singletary, and Aaron D. Ames$^{1}$
\thanks{$^*$ This work was supported by the Air Force Office of Scientific Research.}
\thanks{$^{1}$ The authors are with the California Institute of Technology, 1200 East California Boulevard, Pasadena, CA 91125, USA.
\href{mailto:pakella@caltech.edu}{\texttt{pakella@caltech.edu}},
\href{mailto:urosolia@caltech.edu}{\texttt{urosolia@caltech.edu}},
\href{mailto:asinglet@caltech.edu}{\texttt{asinglet@caltech.edu}},
\href{mailto:ames@caltech.edu}{\texttt{ames@caltech.edu}}}
}

\begin{document}

\maketitle
\thispagestyle{empty}
\pagestyle{empty}

\begin{abstract}
As control systems become increasingly more complex, there exists a pressing need to find systematic ways of verifying them.  To address this concern, there has been significant work in developing test generation schemes for black-box control architectures.  These schemes test a black-box control architecture's ability to satisfy its control objectives, when these objectives are expressed as operational specifications through temporal logic formulae.  Our work extends these prior, model based results by lower bounding the probability by which the black-box system will satisfy its operational specification, when subject to a pre-specified set of environmental phenomena.  We do so by systematically generating tests to minimize a Lipschitz continuous robustness measure for the operational specification.  We demonstrate our method with experimental results, wherein we show that our framework can reasonably lower bound the probability of specification satisfaction.
\end{abstract}

\section{Introduction}
\nocite{video}
An integral aspect of control system design is to ensure that the driven system satisfies some extraneous criteria, \textit{e.g.} safety, robustness to noise/model mismatch, \textit{etc.}  To quantify this assurance, these criteria are oftentimes expressed as temporal logic formulae, and the development of controllers which are guaranteed to satisfy these formulae - termed correct-by-construction controllers - have seen significant interest in the recent past \cite{Nilsson2014,Nilsson2015}.  However, as control systems become increasingly more complex and are subject to more diverse, varied scenarios, developing a correct-by-construction controller becomes progressively more difficult, if not, perhaps, impossible \cite{Seshia2020}.  As a result, the problem remains, as to how to verify that a complex control architecture satisfies the criteria required of it.

In the Test and Evaluation (T\&E) community, significant work has been done to address this verification dilemma.  Specifically, there has been some work to extend traditional safety verification techniques, by iteratively solving for candidate Lyapunov/Barrier certificates based on simulation data/a dynamic model \cite{Prajna2004,Kapinski2014,ames2016control}.  However, the ideal would be to verify the controller's ability to satisfy its criteria despite the presence of confounding environmental phenomena, as these oftentimes lead to a system's inability to satisfy specifications \textit{e.g.} unpredictable human behavior for autonomous cars and adversarial agents in a reconnaissance context~\cite{Seshia2020}.


Numerous, model-based approaches to testing for verification have been studied by the T\&E community \cite{TE_model_based_unadaptive_1,TE_model_based_unadaptive_2,TE_model_based_unadaptive_3}; however, they tend to be sample inefficient.  As a result, there has also been significant work in increasing sample efficiency of these methods~\cite{hansen2016cma,annpureddy2011s}. One such sample efficient method is Bayesian Optimization, which has seen success in terms of fine tuning existing control strategies \cite{calandra2016bayesian, berkenkamp2016bayesian}.  From a testing perspective, identifying environmental phenomena to frustrate satisfaction of the system's operational specifications can also be recast to a minimization problem solvable by Bayesian Optimization~\cite{ghosh2018verifying}. Indeed, there has also been work to identify multiple such phenomena, should they exist \cite{gangopadhyay2019identification}.

\begin{figure}[t]
    \centering
    \includegraphics[width = 0.48 \textwidth]{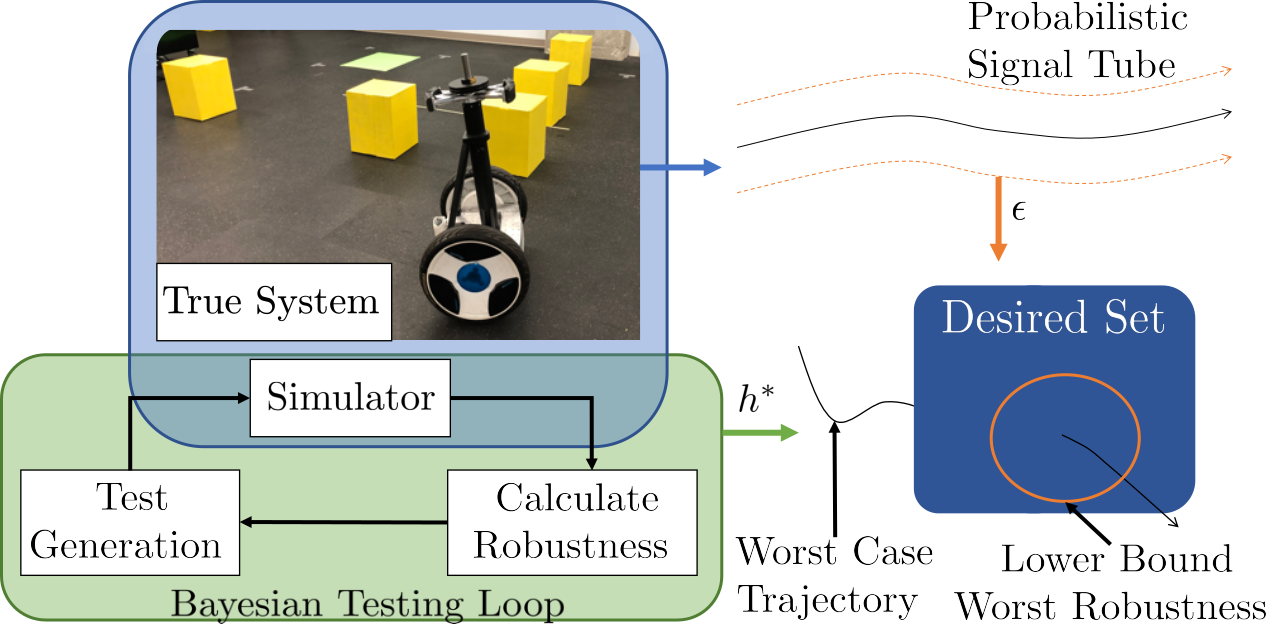}
    \caption{This paper presents a method to lower bound the probability with which a black-box system (the Segway shown top left) satisfies its operational specification (please see Figures~\ref{fig::ros_sim_env} and \ref{fig::test_setup} for the full experimental setup).  This method first determines the worst-case robustness measure, $h^*$, for a system simulator, and uses this $h^*$ and a norm-bound, $\epsilon$, from the signal tube, to lower bound the worst-case robustness for the black-box system, to some minimum probability.
    }
    \label{fig::Title}
    \vspace{-0.2 in}
\end{figure}
\newidea{Our Contribution} We extend prior results in the Bayesian testing community, by lower bounding the probability with which the actual system satisfies its operational specifications.  We do so for Reach-Avoid type STL specifications, as they permit a Lipschitz continuous robustness measure.  As a result, the testing for verification problem can be reformulated as an optimization problem that is solvable by Bayesian Optimization - minimizing this robustness measure subject to an adversarial environment.  Finally, we use the solution to this minimization problem to lower bound the probability by which the true system satisfies the same operational specification.

\newidea{Outline} in Section~\ref{sec::math_background}, we provide a brief mathematical overview of relevant topics.  Section~\ref{sec::prob_setup} formally sets up the problem under study.  Section~\ref{sec::statement_main_result} details the main theorem and its proof, and Section~\ref{sec::propositions} states and proves three propositions generalizing the theorem.  Finally, Section~\ref{sec::experiment} shows how our method provides a reasonable lower bound on the probability that a Segway satisfies its safety specification, by verifying its associated controller.

\section{Problem Formulation}
This section details the necessary mathematical information for the sequel and frames the problem under study.

\subsection{Mathematical Preliminaries}
\label{sec::math_background}
This subsection will be split into three parts encompassing some general notation, a brief description on Signal Temporal Logic, and a brief description of Bayesian Optimization.  

\spacing
\newidea{Notation:} $\mathbb{R}^d$ denotes the $d$-dimensional Euclidean Space. A function, $f:~\mathbb{X}\to\mathbb{Y}$, is $(L,\|\cdot\|_{\mathbb{X}},\|\cdot\|_{\mathbb{Y}})$-Lipschitz continuous, if there exists a positive constant, $L$, such that:
$\|f(x) - f(y)\|_\mathbb{Y} \leq L \|x-y\|_\mathbb{X}~\forall~x,y\in\mathbb{X},
$
where $\|\cdot\|_\mathbb{X}$ and $\|\cdot\|_\mathbb{Y}$ are norms for the normed vector spaces, $\mathbb{X}$ and $\mathbb{Y}$, respectively.  A signal is a function, $s: \mathbb{R}_+ \to \mathbb{X}$, that maps time to a vector in a vector space, $\mathbb{X}$, \textit{i.e.} $s(t) = x \in \mathbb{X}$.  $\mathcal{S}^{\mathbb{X}}_{t_f}$ is the vector space of all signals over the bounded time-frame, $[0,t_f]$, \textit{i.e.} $\mathcal{S}^{\mathbb{X}}_{t_f} = \{ s:\mathbb{R}_+ \to \mathbb{X}~|~\exists~x\in\mathbb{X}~\suchthat~s(t)=x~\forall~t\in[0,t_f]\}$.  If the vector space, $\mathbb{X}$, admits a norm, $\|\cdot\|$, $\|\cdot\|^{t_f} = \max_{t\in[0,t_f]} \|s(t)\|$, is its induced norm over $\mathcal{S}^{\mathbb{X}}_{t_f}$.

\spacing
\newidea{Signal Temporal Logic:} Signal Temporal Logic is a language by which rich, time-varying system behavior can be succinctly expressed.  This language is based on atomic propositions, $\Phi \in \mathcal{A}$, which are boolean valued variables dependent on system behavior:
\begin{equation}
    \Phi(x) = \true \iff x \in \llbracket \Phi \rrbracket = \{x \in \mathbb{R}^n~|~b(x) \sim \mu \}.
\end{equation}
Here, $\mathcal{A}$ is the set of all atomic propositions, $\Phi(x)$ denotes the truth evaluation of $\Phi$ at the state, $x$, $b: \mathbb{R}^n \to \mathbb{R}$, $\mu \in \mathbb{R}$, and $\sim~= \{\geq, \leq, <, >\}$ \cite{Madsen2018,Rizk2011,Gilpin2020}.  Additionally, $\mathcal{A}$ is closed under logical combinations of its components, \textit{i.e.}:
\begin{align}
    \Phi \in \mathcal{A} & \implies \neg \Phi \in \mathcal{A}, \\
    \Phi_1, \Phi_2 \in \mathcal{A} & \implies \Phi_1 \wedge \Phi_2 \in \mathcal{A}~\mathrm{and}~\Phi_1 \lor \Phi_2 \in \mathcal{A}.
\end{align}
Here, $\neg$ denotes negation, $\wedge$ denotes conjunction (and), and $\lor$ denotes disjunction (or).  System specifications, $\psi$, can be defined as follows:
\begin{align}
    \psi & \triangleq \true|\Phi | \neg \psi | \psi_1 \lor \psi_2 | \psi_1 \wedge \psi_2 | \psi_1 \until \psi_2,
\end{align}
where, $\psi_1,\psi_2$ are specifications themselves \cite{baier2008principles}.
Finally, $\mathbb{S}$ is the set of all STL specifications, \textit{i.e.} $\psi \in \mathbb{S}$ \cite{baier2008principles}.

When a signal, $s:\mathbb{R}_+ \to \mathbb{R}^n$, satisfies a specification, $\psi$, by some time, $t$ \textit{i.e.} ensures $\psi = \true$, we write $s(t) \models \psi$.  Here, $\models$ is termed the satisfaction relation, and it is defined as follows:
\begin{align}
    & s(t) \models \Phi \iff \Phi(s(t)) = \true, \\
    & s(t) \models \neg \psi \iff \psi(s(t_f)) = \false, \\
    & s(t) \models \psi_1 \lor \psi_2 \iff s(t) \models \psi_1 \lor s(t) \models \psi_2, \\
    & s(t) \models \psi_1 \wedge \psi_2 \iff s(t) \models \psi_1 \wedge s(t) \models \psi_2, \\
    & s(t) \models \psi_1 \until \psi_2 \iff \exists~t^*\leq t~\suchthat \\
    & \quad \left(s(t') \models \psi_1~\forall~t' < t^*\right) \wedge \left(s(t^*) \models \psi_2 \right).
\end{align}
Finally, for any signal temporal logic specification, $\psi$, evaluated over some bounded time-frame, $[0,t]$, there exists a robustness measure, $\rho$, with which to measure proximity to satisfaction of the specification \cite{Madsen2018}:
\begin{equation}
\rho: \mathcal{S}^{\mathbb{R}^n}_{t} \to \mathbb{R}~\suchthat \rho(s)\geq 0 \iff s(t) \models \psi.\label{eq:robMeasure}
\end{equation}

\newidea{Bayesian Optimization:} Mathematically, Bayesian Optimization is a solution procedure intended to solve optimization problems of the following form:
\begin{equation}
    x^* = \argmin_{x\in \mathrm{A}\subseteq \mathbb{R}^d}c(x),
\end{equation}
where, typically, $d$ is small; $c(x)$ is expensive to evaluate and lacks nice, analytic structure \textit{e.g.} convexity; and A is some set for which membership evaluation is simple \textit{e.g.} a hyperrectangle \cite{Frazier2018}.  

Abstractly, the procedure follows two, main steps, (for a more comprehensive mathematical treatment, please reference \cite{bull2011}).  The first step fits a Gaussian Process to $c$ based on a data-set of sampled values, $\mathcal{D}_N = \{(x_k,y_k = c(x_k)\}_{k=1}^N$ and a set of parameters, $\theta$.  The second step finds the next, sample point via optimizing an acquisition function - in our case, the Expected Improvement function.  Usually, the parameters $\theta$ are updated after each cycle; however, the convergence results of \cite{bull2011} require a static parameterization.  Therefore, we opt for a static parameterization.

\subsection{Problem Setup}
\label{sec::prob_setup}

We consider an uncertain control system as follows:
\begin{equation}\label{eq:uncSys}
    \dot{x} = f(x,u,d,w),
\end{equation}
where the state $x\in\mathcal{X}\subseteq\mathbb{R}^n$, the control input, $u \in \mathcal{U}\subseteq \mathbb{R}^m$, the environmental configuration, $d \in \testspace \subseteq \mathbb{R}^p$, and the disturbance, $w\sim\pi_{\mathrm{env}}$, for the unknown distribution, $\pi_{\mathrm{env}}$.  For this system, we also have a controller,
\begin{equation}\label{eq:policy}
    U:\mathbb{R}^n \times \mathbb{R}^p \rightarrow \mathcal{U}\subseteq \mathbb{R}^n,~\suchthat~u(t)=U(x(t),d).
\end{equation}
Our goal is to certify that the above controller, $U$, which is designed based on a nominal model,
\begin{equation}\label{eq:sim}
    \dot{x} = \hat f(x, u, d),
\end{equation}
satisfies a specification, $\psi$, to some minimum probability, for all environment configurations, $d \in \testspace$, when in closed-loop with the unknown, uncertain system~\eqref{eq:uncSys}.  

For the true and nominal systems from~\eqref{eq:uncSys} and~\eqref{eq:sim}, we define a simulation trajectory, $\hat{\phi}(\cdot)$, and the actual system trajectory it approximates, $\phi(\cdot)$, as follows:
\begin{subequations}
\begin{align*}
    & \hat{\phi}(x_0,u(t),d,t) = x_0 + \int_0^t \hat f(x(s),u(s),d) ds,\\
    & \phi(x_0, u(t), d, w(t), t) = x_0 + \int_0^t f(x(s),u(s),d,w(s)) ds.
\end{align*}
\end{subequations}
\vspace{-0.2 in} \\
As stated, both $\phi(x_0,u(t),d,w(t))$ and $\hat\phi(x_0,u(t),d)$ are signals, under the assumption that neither system ever engenders state values that tend to infinity.

As the simulation model~\eqref{eq:sim} is deterministic, it may be different from the true system~\eqref{eq:uncSys}. Therefore, we introduce the following function to quantify the difference between solutions to the nominal, closed loop system, $\hat{\phi}(x_0,U(x(t),d),d,t)$, and the true, closed loop system, $\phi(x_0,U(x(t),d),d,w(t),t)$:
\begin{align}
 \Delta_\phi(&x_0,U,d,w,t) \label{eq:simError} \\
& \hspace{-0.2 in} = \phi(x_0,U(x(t),d),d,w(t), t) - \hat{\phi}(x_0, U(x(t),d),d, t), \nonumber   
\end{align}
where $w(t) \in \mathcal{W}_t$ is a random signal where, $w(t)\sim\pi_{\mathrm{env}}~\forall~t\in[0,t_f]$, and $\mathcal{W}_{t_f}$ is the set of all random signals over the bounded time-frame, $[0,t_f]$.  Equation~\eqref{eq:simError} permits us to define the accuracy of our nominal closed loop system~\eqref{eq:policy}-\eqref{eq:sim}, over bounded time-frames, $[0,t_f]$:
\simpledef{def::accurate_sim}{The nominal closed loop system~\eqref{eq:policy}-\eqref{eq:sim} is \textit{$(\epsilon, t_f, \lambda,\| \cdot \|)$-accurate}, if and only if
\begin{equation}
\begin{aligned}
\mathbb{P}_w\left[\max_{0\leq t \leq t_f} \left\|\Delta_\phi(x_0,U,d,w(t),t) \right\| \leq \epsilon \right] \geq 1-\lambda,
\end{aligned}
\end{equation}
$\forall~d \in \testspace$ where $\Delta_\phi(x_0,U,d,w)$ is defined in~\eqref{eq:simError}.
}
We require a notion of accuracy for our nominal model, as we will first certify that the nominal closed loop system~\eqref{eq:policy}-\eqref{eq:sim}  satisfies the specification, $\psi$, for all environmental configurations, $d \in \testspace$.  If the nominal closed loop system indeed satisfies $\psi$, then, provided that $\psi$ meets the criteria for the following assumption, we will extend that certificate to the actual system, with some minimum probability:
\simpleassumption{assump::reach_avoid}{The specification, $\psi$, is of the form, $\psi = \true \until \Phi$ or $\psi = \neg (\true \until \neg \Phi)$, where $\Phi \in \mathcal{A}$.}
In effect then, our problem statement is as follows.
\probstate{For a specification, $\psi$, satisfying the criteria for Assumption~\ref{assump::reach_avoid}, identify the minimum probability by which the true closed loop system~\eqref{eq:uncSys}-\eqref{eq:policy} satisfies $\psi$ by some final time, $t_f$, \textit{i.e.} determine $p$ where,
\begin{equation}
\mathbb{P}_w\left[\phi\left(x_0,U(x(t),d),d,w(t),t_f\right) \models \psi \right] \geq p,~\forall~d\in\testspace,   
\end{equation}
}
\section{Main Result}
\label{sec::main_result}
This section is divided into two parts.  First, we state and prove the main result.  Then, we show under which conditions the assumption from our main result are satisfied.

\subsection{Statement of Main Result}
\label{sec::statement_main_result}
When a system's operational specification, $\psi$, satisfies the criteria for Assumption~\ref{assump::reach_avoid}, and the controller~\eqref{eq:policy} is designed based on the nominal model~\eqref{eq:sim}, we devise a method to lower bound the probability by which the true closed loop system~\eqref{eq:uncSys}-\eqref{eq:policy} satisfies $\psi$, for all environmental configurations.  We do so by first assuming that there always exists a Lipschitz robustness measure, $\rho: \mathcal{S}^{\mathbb{R}^n}_{t_f} \to \mathbb{R}$, for specifications, $\psi$, satisfying Assumption~\ref{assump::reach_avoid} (we will prove that such a Lipschitz measure always exists for these specifications, in Section~\ref{sec::propositions}).  Then, we define a nominal, worst-case robustness value, $h^*$, for this robustness measure, $\rho$, and the nominal closed loop system~\eqref{eq:policy}-\eqref{eq:sim}:
\begin{equation}
    \label{eqn::h^*}
    h^* = \min_{d\in\testspace} \rho(\hat{\phi}(x_0,u(t),d)).
\end{equation}
Finally, our result indicates that if $h^*$ is both positive and sufficiently far from zero, then the true closed loop system~\eqref{eq:uncSys}-\eqref{eq:policy} is guaranteed to satisfy $\psi$ to the same minimum probability defined by our simulator~\ref{def::accurate_sim}.
\simpletheorem{thm::main_result}{
Let $\psi$ be a specification satisfying Assumption~\ref{assump::reach_avoid} with an $(L,\|\cdot\|^{t_f},|\cdot|)$-Lipschitz continuous robustness measure, $\rho:\signalspace \to \mathbb{R}$ as defined in~\eqref{eq:robMeasure}.  If the nominal system~\eqref{eq:policy}-\eqref{eq:sim} is $(\epsilon,t_f,\lambda,\|\cdot\|)$-accurate, and the nominal worst-case robustness, $h^*\geq L\epsilon$, then the true system~\eqref{eq:uncSys}-\eqref{eq:policy} satisfies $\psi$ with at least probability $1-\lambda,~\forall~d\in\testspace$, \textit{i.e.}
\begin{equation*}
\mathbb{P}_w\left[\phi\left(x_0,U(x(t),d),d,w(t),t_f\right) \models \psi \right] \geq 1-\lambda,~\forall~d\in\testspace.   
\end{equation*}
}

Theorem~\ref{thm::main_result} states that if the nominal system robustly satisfies $\psi$, $h^* \geq L\epsilon$, and is $(\epsilon, t_f, \lambda, \| \cdot \|)$-accurate, then with probability greater than $1-\lambda$, the true system will satisfy the same specification, $\psi$, as well. \\
\begin{proof}
We start by showing that if $h^* \geq 0$, then,
\begin{equation}
\hat{\phi}(x_0,U(x(t),d),d,t_f) \models \psi~\forall~d\in\testspace.
\end{equation}
This stems directly, as $h^*$ is defined as the minimum value of the robustness measure, $\rho$, over all environmental configurations, $d \in \testspace$. More formally, we have that
\begin{align}
    h^*\geq 0 & \implies & & \hspace{-0.1 in} \rho(\hat{\phi}(x_0,U(x(t),d),d)) \geq 0, ~\forall~d\in\testspace, \\
    & \hspace{0.075 in}\equiv & & \hspace{-0.1 in} \hat{\phi}(x_0,U(x(t),d),d,t_f) \models \psi~\forall~d\in\testspace.
\end{align}
Similar to equation~\eqref{eqn::h^*}, were we to define the true worst-case robustness measure, $H^*$, for the true system~\eqref{eq:uncSys}-\eqref{eq:policy} and some $w(t) \in \mathcal{W}_{t_f}$,
\begin{equation}
    H^* = \min_{d \in \testspace} \rho(\phi(x_0,U(x(t),d),d,w(t))),
\end{equation}
then if $H^* \geq 0$, the true closed loop system~\eqref{eq:uncSys}-\eqref{eq:policy}, is guaranteed to satisfy $\psi$, $\forall~d\in\testspace$ and this, specific $w(t)$.  The remainder of the proof uses $h^*$ to lower bound $H^*$ for all possible $w(t) \in \mathcal{W}_{t_f}$ to at least some minimum probability.  To simplify notation in its presentation, we abbreviate $\hat\phi(x_0,U(x(t),d),d) = \bar\phi$ and $\phi(x_0,U(x(t),d),d,w(t)) = \Tilde\phi$:
\begin{align}
h^* - H^* & = \min_{d \in \testspace} \rho(\bar\phi) - \min_{d \in \testspace} \rho(\Tilde\phi), \\
& = \max_{d \in \testspace} - \rho(\Tilde\phi) - \max_{d\in\testspace}
-\rho(\bar\phi), \\
& \leq \max_{d \in \testspace} \big| \rho(\Tilde\phi) - \rho(\bar\phi) \big|, \\
& \leq L \max_{d\in\testspace} \max_{0\leq t \leq t_f} \|\Tilde{\phi}(t) - \bar\phi(t) \|, \\
& \leq L\epsilon~\mathrm{with~at~least~probability~} 1-\lambda.
\end{align}
Note that the second to last line is valid $\forall~w(t)\in\mathcal{W}_{t_f}$.  The transition to the last line arises as the nominal system is $(\epsilon,t_f,\lambda,\|\cdot\|)$-accurate.  Furthermore, this probability is over the random signals, $w(t) \in \mathcal{W}_{t_f}$, which is why the random signal, $w(t)$, does not appear in the last line.  Hence,
\begin{align}
    h^* \geq  L\epsilon & \implies & & \hspace{-0.15 in} \mathbb{P}_w[H^* \geq 0] \geq 1-\lambda~\forall~d\in\testspace, \\
    & \hspace{0.08 in} \equiv & & \hspace{-0.15 in} \mathbb{P}_w\left[\Tilde \phi(t_f) \models \psi \right] \geq 1-\lambda~\forall~d\in\testspace,
\end{align}
where we abbreviate $\phi(x_0,U(x(t),d),d,w(t)) = \Tilde\phi$.
\end{proof}

\subsection{Extension of Main Result}
\label{sec::propositions}
Theorem~\ref{thm::main_result} assumes existence of an $(L,\|\cdot\|^{t_f},|\cdot|)$-Lipschitz continuous robustness measure, $\rho$, for specifications, $\psi$, satisfying Assumption~\ref{assump::reach_avoid}.  While this seems restrictive, the following three propositions show that such a Lipschitz robustness measure, $\rho$, always exists for specifications, $\psi$, satisfying Assumption~\ref{assump::reach_avoid}.

The first proposition develops a Lipschitz continuous function, $h$, with which to measure satisfaction of any predicate, $\Phi \in \mathcal{A}$:
\simpleproposition{proposition::spec_lipschitz}{Let $\Phi \in \mathcal{A}$.  For some norm, $\|\cdot\|$ on $\mathbb{R}^n$, there exists an $(L,\|\cdot\|,|\cdot|)$-Lipschitz function, $h:\mathbb{R}^n \to \mathbb{R}$, such that $h(x)\geq 0 \iff \Phi(x) = \true$.
}
\begin{proof}
For $\Phi = \true$, any constant, positive function $h$ suffices \textit{e.g.} $h(x) = 1$; likewise, for $\Phi = \false$, any constant, negative function $h$ suffices.  It remains to show the same result for each $\Phi \in \mathcal{A}\setminus\{\true,\false\}$.  Here, we note that as $\Phi \neq \true,\false$, its associated truth region, $\llbracket \Phi \rrbracket$, has a non-trivial boundary, \textit{i.e.} $\partial \llbracket \Phi \rrbracket \neq \varnothing$.  As a result, we can define a signed, distance function, $h$,
\[
l(x) = \min_{y\in\partial\llbracket \Phi \rrbracket} \|x-y\|, \quad
h(x) = 
\begin{cases}
l(x) & x \in \llbracket \Phi \rrbracket, \\
-l(x) & x \in \llbracket \neg \Phi \rrbracket.
\end{cases}
\]
By definition, $h$ is positive if and only if $\Phi = \true$.  Also, $h$ carries the same Lipschitz constant, $L=1$, as the set-distance function, $l$, thus completing the proof.
\end{proof}

Similar to Proposition~\ref{proposition::spec_lipschitz}, the second proposition develops a Lipschitz robustness measure, $\rho$, for any specification, $\psi$, that satisfies Assumption~\ref{assump::reach_avoid}:
\simpleproposition{proposition::lipschitz_robustness}{
Let $\psi$ be a specification that satisfies Assumption~\ref{assump::reach_avoid}, with $\Phi \in \mathcal{A}$.  Assume that there exists an $(L,\|\cdot\|,|\cdot|)$-Lipschitz function, $h$, such that $h(x) \geq 0$ if and only if $\Phi(x) =\true$. Then for $\psi$, there also exists an $(L,\|\cdot\|^{t_f},|\cdot|)$-Lipschitz robustness measure, $\rho: \signalspace \to \mathbb{R}$, as in~\eqref{eq:robMeasure}, \textit{i.e.},
\begin{equation}
    |\rho(s') - \rho(s)| \leq L \max_{0 \leq t \leq t_f} \|s'(t) - s(t) \| = L\|s'-s\|^{t_f}.
\end{equation}
}
\begin{proof}
It is known from prior works, \textit{e.g.} \cite{Madsen2018}, that the following are valid robustness measures for specifications, $\psi$, satisfying Assumption~\ref{assump::reach_avoid}, as $h(x) \geq 0 \iff \Phi(x) = \true$:
\[
\rho(s) = 
\begin{cases}
\max\limits_{0\leq t \leq t_f} h(s(t)), & \mbox{If}~\psi=\true\until\Phi, \\
\min\limits_{0\leq t \leq t_f} h(s(t)), & \mbox{If}~\psi=\neg(\true\until\neg\Phi).
\end{cases}
\]
We will show that the first measure is Lipschitz.
\begin{align}
    |\rho(s') - \rho(s)| & = \left|\max_{0\leq t \leq t_f} h(s'(t)) - \max_{0\leq t \leq t_f} h(s(t)) \right|, \\
    & \leq \max_{0 \leq t \leq t_f} \left|h(s'(t)) - h(s(t)) \right|, \\
    & \leq L \max_{0 \leq t \leq t_f} \|s'(t) - s(t) \|.
\end{align}
Following a similar chain of logic shows that the second measure is Lipschitz, thus completing the proof.
\end{proof}

Finally, Propositions~\ref{proposition::spec_lipschitz} and \ref{proposition::lipschitz_robustness} ensure that there always exists a Lipschitz continuous robustness measure, $\rho$, for any specification, $\psi$, satisfying the conditions for Assumption~\ref{assump::reach_avoid}.
\simpleproposition{proposition::robustness_guaranteed}{
For any specification, $\psi$, that satisfies Assumption~\ref{assump::reach_avoid}, there always exists a Lipschitz continuous robustness measure, $\rho$, of the form in~\eqref{eq:robMeasure}.
}
\begin{proof}
The proof is a direct application of Propositions~\ref{proposition::spec_lipschitz} and \ref{proposition::lipschitz_robustness}.  Proposition~\ref{proposition::spec_lipschitz} guarantees the existence of at least one $(L,\|\cdot\|,|\cdot|)$-Lipschitz function, $h$, for $\Phi \in \mathcal{A}$, $L\in\mathbb{R}_+$, and norm, $\|\cdot\|$, on $\mathbb{R}^n$.  Using this function, $h$, in the associated robustness measure, $\rho$, for $\psi$, as specified in Proposition~\ref{proposition::lipschitz_robustness}, guarantees the existence of at least one Lipschitz robustness~measure.
\end{proof}

As a result, Proposition~\ref{proposition::robustness_guaranteed} indicates that, for any specification, $\psi$, that satisfies Assumption~\ref{assump::reach_avoid}, there always exists a Lipschitz continuous robustness measure, $\rho$.  As a result, if the nominal system is $(\epsilon, t_f, \lambda, \|\cdot\|)$-accurate, with respect to the same norm, $\|\cdot\|$, that induces the norm for which $\rho$ is Lipschitz, $\|\cdot\|^{t_f}$, the results of Theorem~\ref{thm::main_result} still hold.

\section{Experimental Results}
\label{sec::experiment}
\begin{figure}[t]
    \centering
    \includegraphics[width = 0.47\textwidth]{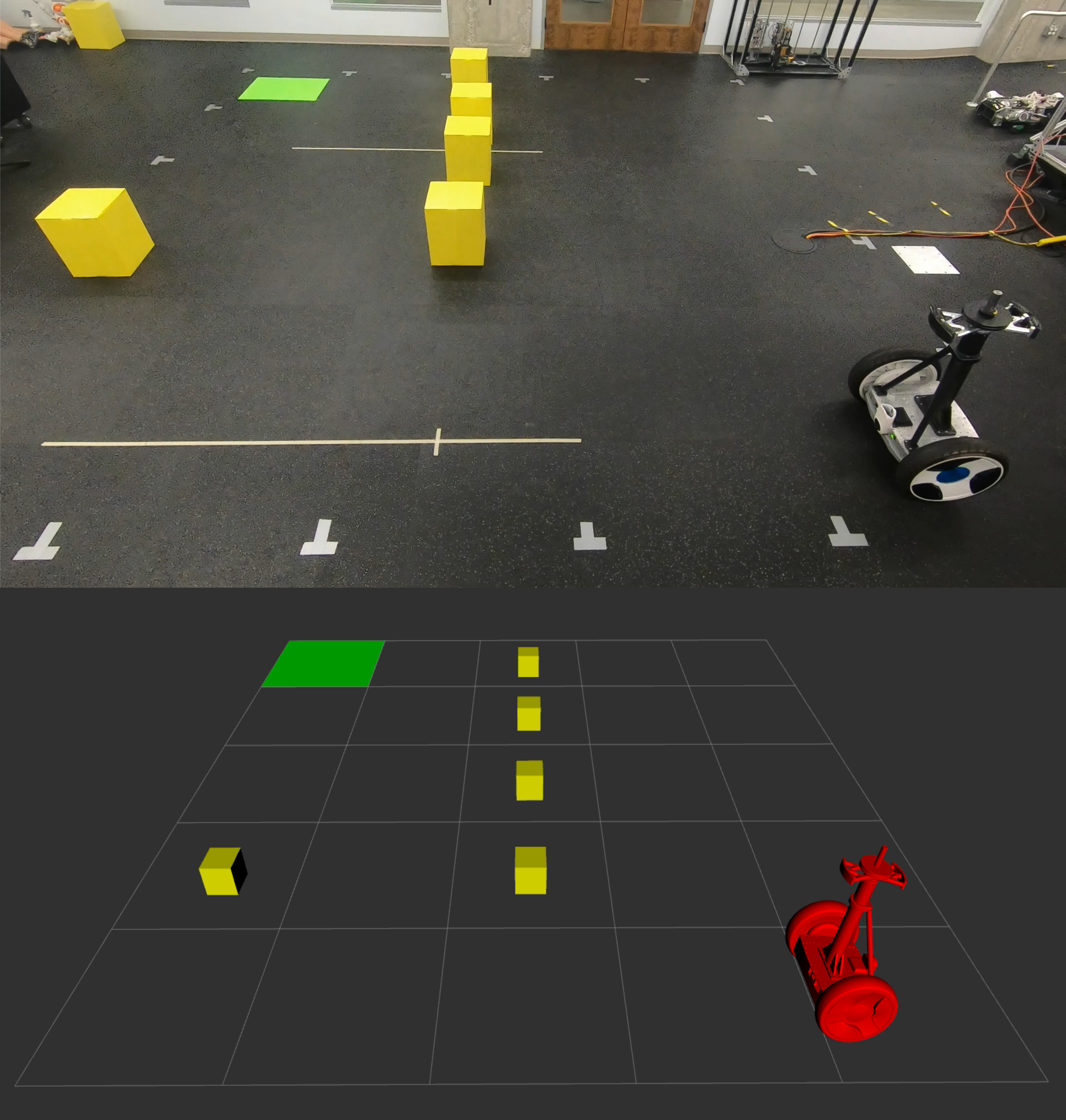}
    \caption{A picture of the (bottom) ROS-based simulation environment  we used to approximate our (top) true, Segway system.}
    \label{fig::ros_sim_env}
    \vspace{-0.2 in}
\end{figure}

Our goal is to verify that a Nonlinear MPC controller, designed to steer a Segway to a goal while avoiding obstacles, satisfies a safety specification, $\psi$, incumbent on the Segway.  This verification question arose when, under normal operation, this controller appeared to successfully steer the Segway to the goal while also satisfying $\psi$.  Hence, we want to use Theorem~\ref{thm::main_result}, to verify that this controller always satisfies $\psi$, regardless of the goals provided to the Segway.  The experimental setup and accompanying simulation, for this verification procedure, is shown in Figure~\ref{fig::ros_sim_env}.

\begin{figure}[t]
    \centering
    \includegraphics[width = 0.48\textwidth]{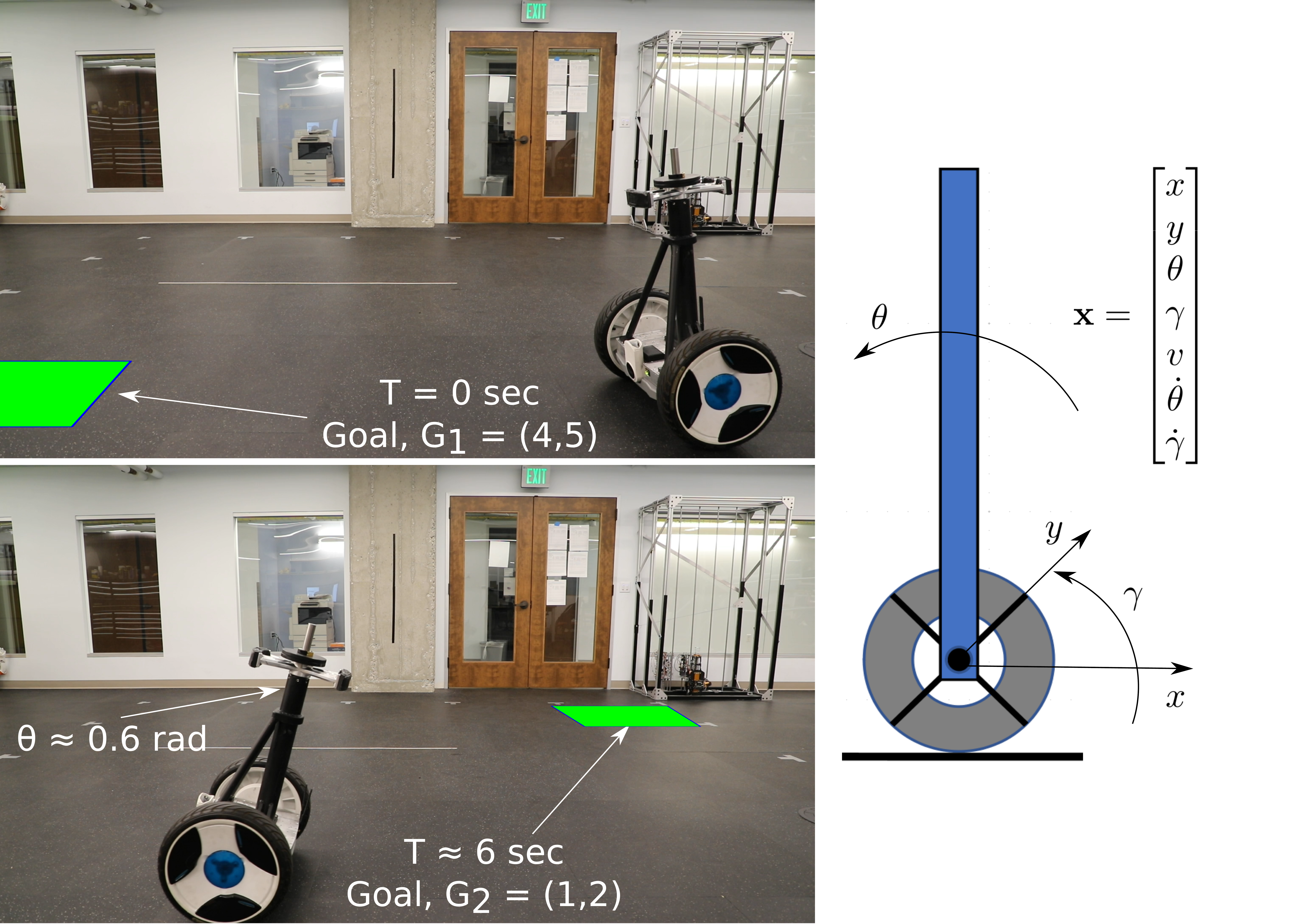}
    \caption{Shown above is the Segway operating with an example environmental configuration, $d = [4,5,1,2,4.885]^T$.  Each vector, $d = [G_1,G_2,T]^T \in \mathbb{R}^5$ denotes the first goal cell to which the Segway is to navigate, $G_1 \in \{1,2,3,4,5\}^2$, the second goal cell, $G_2 \in \{1,2,3,4,5\}^2$, and the time, $T \in [0,10]$, the Segway is to switch between going to $G_1$ and going to $G_2$.  The example shown is a worst-case configuration, the experimental trajectory for which is shown, in red, in Figure~\ref{fig::robustness_bounding}.}
    \label{fig::test_setup}
    \vspace{-0.2 in}
\end{figure}

To use Theorem~\ref{thm::main_result} for this verification, we require a feasible space of environmental configurations, $\testspace$.  As we are trying to verify the controller's capacity to satisfy $\psi$ regardless of the goals provided to it, we specify a vector, $d\in\testspace \subset \mathbb{R}^5$, to be a vector of two goals, $G_1,G_2$, and a switching time, $T$.  This time, $T$, determines when the Segway switches from navigating to the first goal, $G_1$, to navigating to the second goal, $G_2$.  Goals are represented as shaded, cellular regions, as in the green region in Figure~\ref{fig::ros_sim_env}.  Hence, $G_1,G_2 \in \{1,2,3,4,5\}^2$ are cells in this $5\times 5$ grid, and the switching time, $T \in [0,10]$. Figure~\ref{fig::test_setup} portrays a graphical representation of the Segway undergoing a specific environmental configuration, and the full, mathematical setup~is:
\begin{align*}
    & x = [x,y,\theta,\psi,v,\dot{\theta},\dot\psi] \in \mathbb{R}^7,~
    d = [G_1, G_2, T]^T \in \testspace \subset \mathbb{R}^5,\\
    & \llbracket \Phi \rrbracket = \{\theta\in\mathbb{R}~|~|\theta| \leq 0.7~\mathrm{rad}\},~
    \psi = \neg(\true \until \neg \Phi).
\end{align*}
Given the above environment and safety specification, $\psi$, the measurement function $h$ and the robustness measure are defined as:
\begin{align}
        h(x) & = 0.7 - |\theta|, \label{eqn::experiment_h}\\
    \rho(\bar\phi) & = \min\limits_{0\leq t \leq t_f} h(\bar\phi(t)) \label{eqn::experiment_rho}.
\end{align}
Here, we abbreviate $\hat\phi(x_0,U(x(t),d),d) = \bar \phi$.  With this setup, Theorem~\ref{thm::main_result} requires the parameters, $\epsilon, t_f$, and $\lambda$, for our ROS-based simulator, and the nominal worst-case robustness $h^*$ as in~\eqref{eqn::h^*}, for the robustness measure~\eqref{eqn::experiment_rho}.

\begin{figure}[t]
    \centering
    \includegraphics[trim = 5mm 2mm 0mm 0mm, width = 0.48\textwidth]{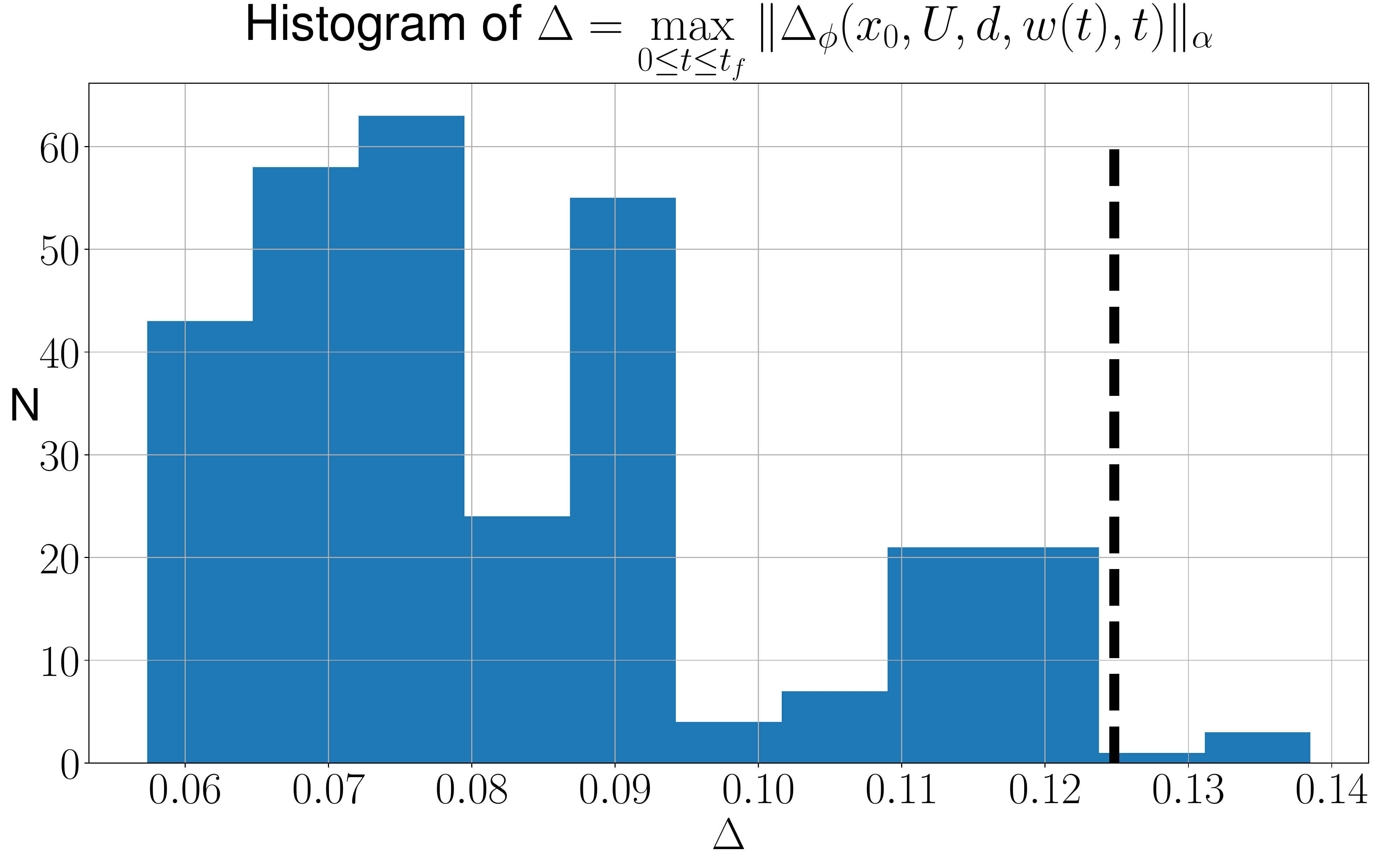}
    \caption{Shown above is a histogram of the sampled, upper-bound on angular trajectory deviation, $\Delta$, with respect to the norm, $\|\cdot\|_\alpha$.  The dashed, black line indicates the cutoff $\epsilon$ value, $\epsilon = 0.125$, for which $\mathbb{P}[\Delta \leq \epsilon] \geq 0.95$.}
    \label{fig::epsilon_vals}
    \vspace{-0.2 in}
\end{figure}

To determine the parameters for our simulator, we note that our measurement function~\eqref{eqn::experiment_h} is $(1,\|\cdot\|_\alpha,|\cdot|)$-Lipschitz with respect to the norm,
\[
\|x\|_\alpha = \sum_{i=1}^7~\alpha_ix_i^2,\quad
\alpha_i = \begin{cases}
1 & \mbox{If }i = 3 \\
10^{-6} & \mathrm{Else}.
\end{cases}
\label{eqn::norm_alpha}
\]
Hence, robustness measure~\eqref{eqn::experiment_rho} is also $(1,\|\cdot\|_\alpha^{t_f},|\cdot|)$-Lipschitz.  As Theorem~\ref{thm::main_result} requires that our simulator be accurate with respect to the same norm, $\|\cdot\|_\alpha$, that induces the norm over which $\rho$ is Lipschitz, $\|\cdot\|_\alpha^{t_f}$, we determined the accuracy of our simulator with respect to this norm, $\|\cdot\|_\alpha$.  To do so, we recorded the maximum simulation error, $\Delta = \max_{0\leq t \leq t_f} \|\Delta_\phi(x_0,U,d,w(t),t)\|_\alpha$~\eqref{eq:simError}, for $N=300$ different pairs of simulation and system trajectories, wherein the starting, goal, and obstacle locations were the same for each pair.  The recorded data is shown in Figure~\ref{fig::epsilon_vals}.  Here, we inherently assume that the maximum norm deviance, $\Delta$, is a random variable distributed by some unknown distribution, $\pi_\Delta$.  As we
do not have access to $\pi_\Delta$, we construct a Monte-Carlo based estimate probability distribution, $\pi^N_\Delta$. This distribution, $\pi^N_\Delta$, is known to be an unbiased estimator of $\pi_\Delta$, for which the absolute value of the difference in variance of the two probability measures is bounded above by $\frac{1}{N}$.  Based on this estimate distribution, $\pi^N_\Delta$, our data indicates that our ROS-based simulator is $(\epsilon = 0.125, t_f = 10, \lambda = 0.05, \|\cdot\|_\alpha)$-accurate, as $\mathbb{P}[\Delta \leq \epsilon = 0.125] \geq (1-\lambda = 0.95)$.

\begin{figure}[t]
    \centering
    \includegraphics[trim = 3mm 0mm 0mm 0mm, width = 0.48\textwidth]{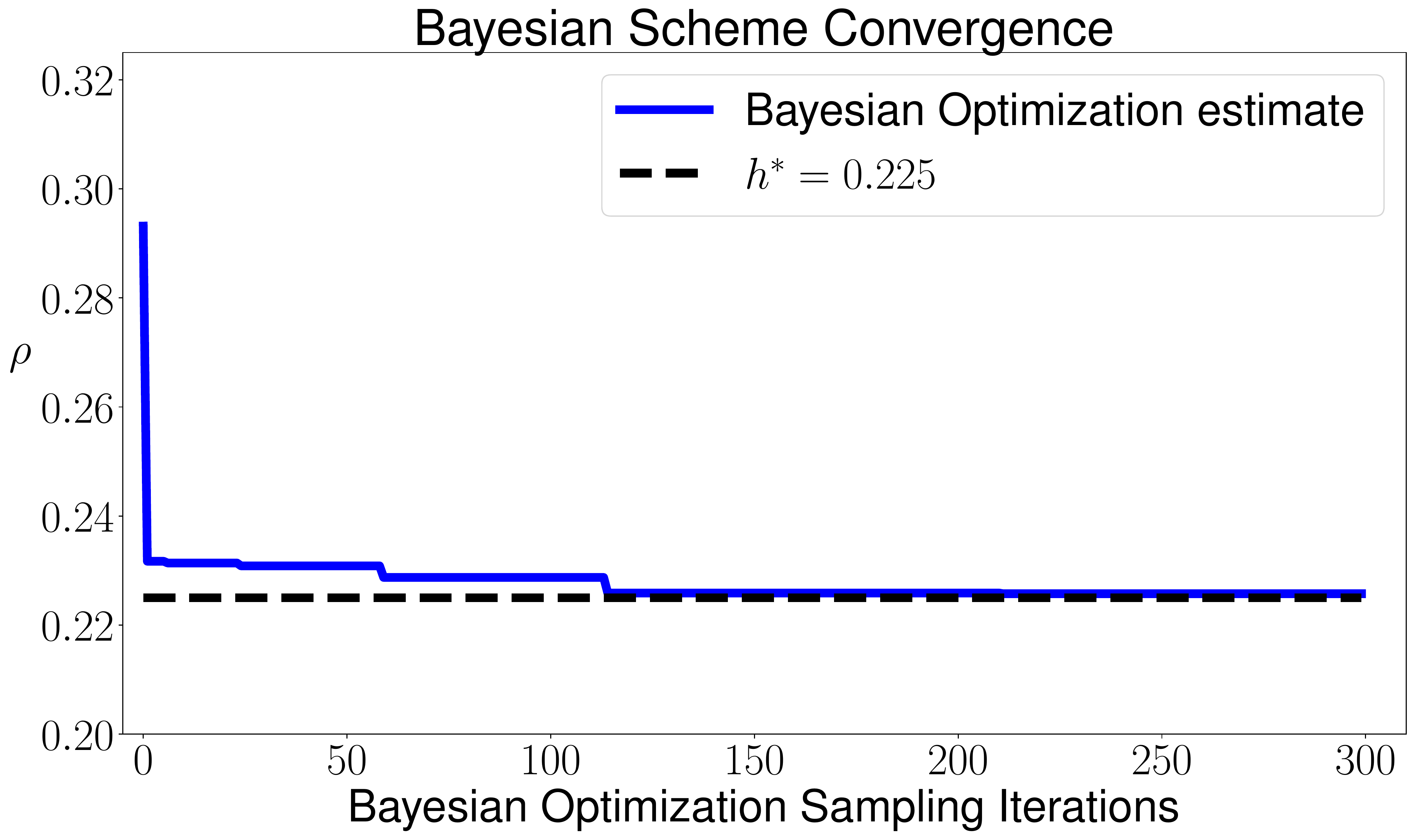}
    \caption{Shown above is the convergence of the Bayesian-Testing framework, minimizing the robustness measure, $\rho$, stated in Section~\ref{sec::experiment}, over the ROS-based simulation of our Segway.  Notice that by 125 iterations, the Bayesian optimization estimate converged to $h^* = 0.225$.}
    \label{fig::convergence}
    \vspace{-0.2 in}
\end{figure}

It remains to determine, for this simulator, the nominal worst-case robustness measure, $h^*$, defined in equation~\eqref{eqn::h^*}, with respect to the robustness measure~\eqref{eqn::experiment_rho}.  To solve optimization problem~\eqref{eqn::h^*}, we used the Bayesian optimization scheme detailed in \cite{bull2011}, which was proven to converge in expectation.  Figure~\ref{fig::convergence} shows the convergence results of this algorithm, when run for $300$ iterations, in attempting to solve optimization problem~\eqref{eqn::h^*}. By roughly 125 iterations, the algorithm had converged to an $h^* = 0.225$, which, based on the convergence results of this algorithm in \cite{bull2011}, we assume to be the solution to optimization problem~\eqref{eqn::h^*}.  Furthermore, this $h^*$ arose for the environmental configuration, $d=[G_1 = (4,5), G_2 = (1,2), T = 4.885]^T$.

\begin{figure}[t!]
    \centering
    \includegraphics[trim = 0mm 2mm 0mm 0mm, width = 0.48\textwidth]{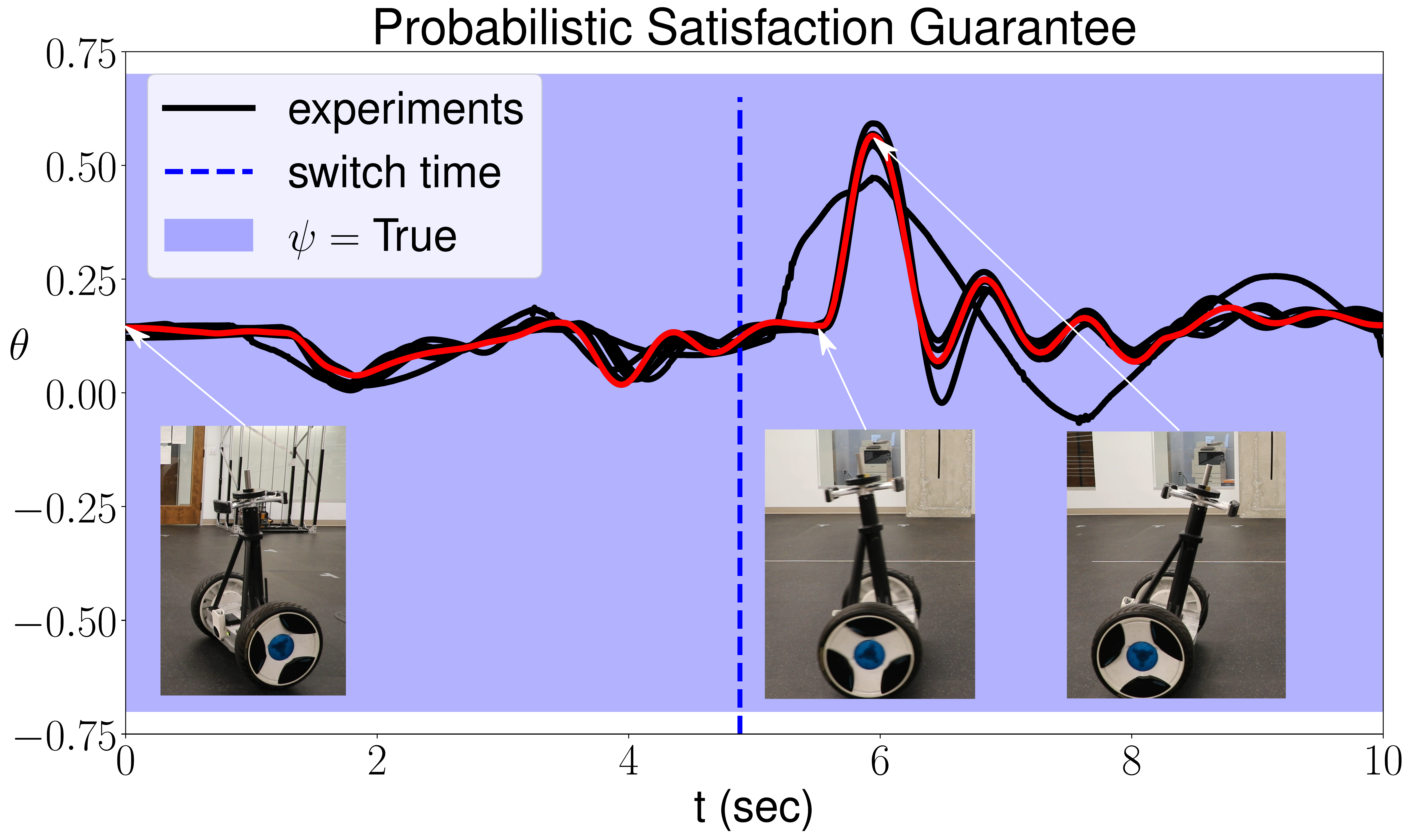}
    \caption{Shown above is the angular trajectories of the Segway undergoing seven independent trials of the worst case environmental configuration derived in Section~\ref{sec::experiment}.  The goal switching time, $T$, is demarcated by the dashed, blue line.  Several snapshots of the Segway, during the trajectory highlighted in red, are shown at the bottom.  We anticipate the system to lie within the lavender region shown, with probability $\geq 0.95$, and as such, satisfy its safety specification, $\psi$, to the same minimum probability. }
    \label{fig::robustness_bounding}
    \vspace{-0.2 in}
\end{figure}

As a result, Theorem~\ref{thm::main_result} indicates that, as $(h^* = 0.225) \geq (L\epsilon = 0.125)$, 
\begin{equation}
    \label{eqn::probabilistic_verification}
    \mathbb{P}_w\left[\Tilde\phi(t_f) \models \psi \right] \geq 0.95~\forall~d\in\testspace.
\end{equation}
Here, we abbreviated $\phi(x_0,U(x(t),d),d,w(t)) = \Tilde\phi$.  To check the accuracy of this probabilistic claim, seven, independent runs of the Segway undergoing this worst-case environment configuration, $d = [4,5,1,2,4.885]$, are shown in Figure~\ref{fig::robustness_bounding}.  The probabilistic verification claim~\eqref{eqn::probabilistic_verification} indicates that, with probability at least $0.95$, experimental trajectories should satisfy $\psi$, as they should all lie within the lavender box - and they indeed all lie within that box.  For a video summarizing this paper, in addition to experimental demosntrations, please reference \cite{video}.
\section{Conclusion}
Our main contribution extends prior work done in the Bayesian testing community, by describing a method to lower bound the probability by which a system will satisfy an operational specification in practice.  Our method constructs a probabilistic signal tube around a simulation trajectory, in which
we anticipate the real system trajectory to lie, with some minimum probability.  Then, we determine the worst-case robustness measure for this nominal system, which, in conjunction with the signal tube, we use to lower bound the worst case robustness measure for the true system.  If this lower bound is positive, the controller is verified to satisfy its specification, with the same minimum probability.

\renewcommand{\baselinestretch}{0.98}
\bibliographystyle{ieeetr}
\bibliography{bibliography/collected_works} 
\vspace{-2 in}

\end{document}